\documentclass[aps pra,floatfix,twocolumn,showpacs, superscriptaddress, amsmath,amssymb,tightenlines,epsfig]{revtex4-1}
\usepackage{graphicx}
\usepackage{dcolumn}	
\usepackage{bm}			
\usepackage{amsfonts}
\usepackage{xspace}
\usepackage{color}
\usepackage{epstopdf}
\usepackage{array}
\usepackage{color}
\definecolor{plum}{rgb}{0.5,0.01,0.5}

\begin{document}

\title{Coherence and linewidth of a continuously pumped atom laser at finite temperature}
\author{Geoffrey M. Lee}
\affiliation{The University of Queensland, School of Mathematics and Physics, Qld 4072, Australia}
\author{Simon A. Haine}
\email{haine@physics.uq.edu.au}
\affiliation{The University of Queensland, School of Mathematics and Physics, Qld 4072, Australia}
\author{Ashton S. Bradley}
\affiliation{Jack Dodd Centre for Quantum Technology, Department of Physics,
University of Otago, PO Box 56, Dunedin, New Zealand}
\author{Matthew J. Davis}
\affiliation{The University of Queensland, School of Mathematics and Physics, Qld 4072, Australia}
\affiliation{JILA, University of Colorado, 440 UCB, Boulder, Colorado 80309, USA}

\begin{abstract}
A continuous wave atom laser formed by the outcoupling of atoms from a trapped Bose-Einstein condensate (BEC) potentially has  a range of  metrological applications.  However, in order for the device to be truly continuous, a mechanism to replenish the atoms in the BEC is required. Here we calculate the temporal coherence properties of a continuously pumped atom laser beam outcoupled from a trapped Bose-Einstein condensate which is replenished from a reservoir at finite temperature.  We find that the thermal fluctuations of the condensate can significantly decrease the temporal coherence of the output beam due to atomic interactions between the trapped BEC and the beam, and this can impact the metrological usefulness of the device.  We demonstrate that a Raman outcoupling scheme imparting a sufficient momentum kick to the atom laser beam can lead to a significantly reduced linewidth.
\end{abstract}
\maketitle

\section{Introduction}

A rudimentary atom laser can be realised by outcoupling atoms from a trapped Bose-Einstein condensate. This has previously been achieved using either radio-frequency (RF)  \cite{Mewes1997,Bloch1999,Coq2001,Robins2004} or Raman transitions \cite{Hagley1999,Robins2006} to transfer the BEC atoms to untrapped states which then propagate freely.  The majority of experiments performed to date have outcoupled atoms from a BEC  that is not replenished, leading to an overall decrease of beam flux as time proceeds.  This places an upper limit on the integration time for any measurement  \cite{Lahaye2005}.  A truly continuous wave (CW) atom laser  will not only extend the  integration time of any measurement, but has other potential advantages such as a reduction in the linewidth of the beam due to gain narrowing \cite{Hope2000}.  However, this will require a method of replenishing the number of atoms in the BEC so that the system achieves a steady state output.  Continuous atom lasers have the potential for applications in metrology, interferometry and precision measurement  \cite{Gustavson1997,Shin2004}, and the improved first-order coherence properties of BECs \cite{Bloch2000,Esslinger2000,Ritter2007} may provide a number of advantages over thermal atom sources \cite{szigeti2012} currently used in, for example, Sagnac interferometry for the detection of rotations \cite{Gustavson1997,Durfee2006,Eckert2006}.

The first experiment that demonstrated a continuous source of Bose-condensed atoms was performed by Chikkatur \emph{et al.}~\cite{Chikkatur2002}. This was achieved by periodically replenishing a BEC held in an optical dipole trap with a new condensate transferred using optical tweezers.  While the team was able to demonstrate a BEC of more than $10^6$ atoms at all times, this approach has the limitation that the coherence of the  BEC is disturbed every time the condensates are combined due to their independent phases.  An alternate approach was demonstrated by Lahaye \emph{et al.}~\cite{Lahaye2005} who performed evaporative cooling on a magnetically guided continuous atomic beam. They observed a significant increase in phase-space density, but did not reach Bose-Einstein condensation.  More recently Robins \emph{et al.}\ have demonstrated a pumped atom laser whereby atoms from one condensate acting as a source undergo a stimulated transition into a second  condensate from which an atom laser beam was extracted using RF outcoupling \cite{Robins2008}.  Recently a continuous strontium BEC formed by laser cooling has been demonstrated~\cite{stellmer_pasquiou_13}, which is potentially an excellent source for an atom laser.  While these experiments have demonstrated the separate elements necessary for a continuous atom laser, they have yet to be combined in a single experiment.

Many theoretical proposals for continuously-pumped operation have been made, using either evaporative cooling \cite{Wiseman1996,Holland1996a,dennis2012} or spontaneous emission \cite{Bhongale2000,Castin1998,Wiseman1995},  A number of studies have already considered the properties of a CW atom laser at zero temperature \cite{Edwards1999,Schneider1999,Naraschewski1997,Hope2000a,Haine2002, Haine2003, Haine2004, Johnsson2005, Johnsson2007,Johnsson2007_1, Szigeti2010}.  However, one of the most likely experimental routes to a true CW atom laser will involve replenishing the BEC via cooling from a reservoir containing thermal atoms so that the condensate mode is maintained in a steady state via Bose-stimulated collisions, similar to the experiment of Stellmer~\emph{et al.}~\cite{stellmer_pasquiou_13}.
This will avoid the deleterious linewidth and heating effects associated with the merging of initially separated condensates with independent phases \cite{Chikkatur2002}.  However, this will also mean that a CW atom laser will necessarily operate at finite temperature, and hence the effect of thermal fluctuations on the atom laser coherence should be accounted for.  An initial study of such issues has been performed by Proukakis \cite{Proukakis2003} using a classical field model for the BEC replenished by an undepleted thermal reservoir. Here we build on this work by implementing a conceptually similar model but with specific improvements.  In particular, we implement an appropriate energy cutoff in the harmonic oscillator basis for the classical field, and propagate the atom laser beam in a plane wave basis well beyond the influence of the trapped BEC, as illustrated in Fig.~\ref{fig:temptrace}.  We focus on understanding the linewidth of the atom laser beam as a function of both the temperature and the momentum kick of the Raman outcoupling.

\section{Model of a continuously pumped atom laser} 
We model a trapped BEC at finite temperature using the simple growth theory of the Stochastic Projected Gross-Pitaevskii equation (SPGPE) formalism \cite{Blakie2008,Gardiner2002,Gardiner2003}, in which the energy damping of the scattering terms are neglected.  This has previously been successfully applied to model the formation of Bose-Einstein condensates from evaporative cooling~\cite{weiler_neely_08}.
 Briefly, the second quantised Bose-field operator is divided at a cutoff energy $E_{\rm cut}$ into a high occupation ($N_k \gg 1$) coherent region, in which interaction effects and thermal fluctuations are significant, and a low occupation incoherent region that can be treated as a thermal gas.    The dynamics of the entire coherent region can then be treated approximately as a classical field.  For the problem to be tractable we treat the incoherent region as a undepleted reservoir at fixed temperature $T$ and chemical potential $\mu$.  We expect this is a reasonable approximation in steady state if we imagine that we can replenish thermal atoms at a constant rate while undergoing continuous  cooling, e.g.~\cite{stellmer_pasquiou_13,dennis2012}.  The stationary incoherent region thus continuously replenishes the coherent region through Bose-stimulated collisions.  The atom laser beam is formed by implementing Raman outcoupling from the coherent region, which depletes not only the condensate but also the non-condensed atoms, resulting in a thermally broadened linewidth for the atom laser beam.

We consider a magnetically trapped $^{87}$Rb condensate containing approximately $10^5$ atoms. We outcouple atoms into an atomic waveguide, as demonstrated by Guerin \emph{et al.}~\cite{Guerin2006}, by transferring them to a magnetically insensitive state via a Raman transition which gives the outcoupled atoms a momentum kick in the positive $x$ direction. The waveguide provides confinement in the radial direction. We choose experimentally reasonable trapping frequencies of $(\omega_x,\omega_\perp)$ = $2\pi\times(10,200)$ Hz. For relatively tight transverse confinement, we  neglect the dynamics in the radial direction and simulate the system in one dimension. 

The equations of motion for the trapped field $\psi(x,t)$ and outcoupled atoms $\phi(x,t)$ are
\begin{eqnarray}
\label{eqn:trapped}
d \psi(x) &=&-\frac{i}{\hbar}\mathcal{P} \left[(\hat{L} + g_{12} |\phi(x)|^2|)\psi(x)
-\hbar\Omega e^{-ik_0x}\phi(x)\right] dt \nonumber\\
 &+& \mathcal{P}\left[\gamma(\mu-\hat{L})  \psi(x)dt+ dW_\gamma(x,t)\right],\\
d \phi(x) &=&-\frac{i}{\hbar}\bigg[ \left( \frac{-\hbar^2}{2m}\frac{\partial^2 }{\partial x^2}
 +g_{12} |\psi(x)|^2 + g_{22}|\phi(x)|^2  - \hbar \delta \right) \phi(x) \nonumber \\ 
 &-& \hbar\Omega e^{ik_0x}\psi(x) \bigg] dt \, ,
\end{eqnarray}
with 
\begin{equation}
\label{eqn:gpoperator1d}
 \hat{L}=-\frac{\hbar^2}{2m}\frac{\partial^2}{\partial x^2} + \frac{1}{2}m\omega_x^2 x^2+g_{11} |\psi(x)|^2 \, ,
\end{equation}
where $m$ is the atomic mass, and $\mu$ and $T$ are the chemical potential and temperature of the reservoir.  The recoil momentum and Rabi frequency of the Raman transition are $\hbar k_0$ and $\Omega$ respectively, with a two-photon detuning of $\delta$. 
The operator ${\mathcal{P}}$ projects $\psi(x)$ onto the truncated basis of harmonic oscillator eigenstates with energy $\epsilon \le E_{\rm cut}$.  The effective 1D interaction strengths are $g_{ij} = \frac{4\pi\hbar^2 a_{ij}}{2m\sigma_{\perp}}$,  where $\sigma_\perp$ is determined by integrating out the transverse dimension using a Gaussian variational  anstaz, and $a_{ij}$ is the $s$-wave scattering length for collisions between states $i$ and $j$. In this work we choose $a_{11} = a_{12} = a_{22} = 5.29 $ nm, which is approximately true for the $|F=1, m_F = -1\rangle$ and $|F=2, m_F=0\rangle$ states of $^{87}$Rb \cite{egorov2013}. The noise is complex Gaussian and has a non-vanishing correlator $
\langle dW^*_\gamma(x,t) dW_\gamma(x',t) \rangle = 2 \gamma \delta(x-x') dt$.
The quantity $\gamma$ is the rate constant for the exchange of particles between the thermal reservoir and the classical field, for which we use the estimate $ \gamma= 12 m a_{11}^2 k_B T/\pi\hbar^3$ \cite{Bradley2008}. The steady state of the atom laser is independent of this quantity inside the weak outcoupling regime such that the condensate is not significantly depleted. We chose  $\Omega = 1$ Hz for all our simulations in order to stay in the weak outcoupling regime~\cite{Robins2005}. 

We simulate the classical field $\psi(x)$ using the harmonic oscillator basis as described in Ref.~\cite{Blakie2005,Blakie2008}, and present results for a cutoff of $E_{\rm cut} = 150 \hbar \omega_x$.  We have increased the cutoff to $E_{\rm cut} = 300 \hbar \omega_x$ in other simulations and found no significant difference in our results.
The beam $\phi(x)$ is simulated on a rectangular grid, and we make use of numerical grid interpolation for the coupling terms.  Classical field atoms are transferred to the beam with a momentum $\hbar k_0$, and we implement an imaginary absorbing potential at the edge of the atom laser beam grid.

\begin{figure}
	\centering
	\includegraphics[width=0.85\columnwidth]{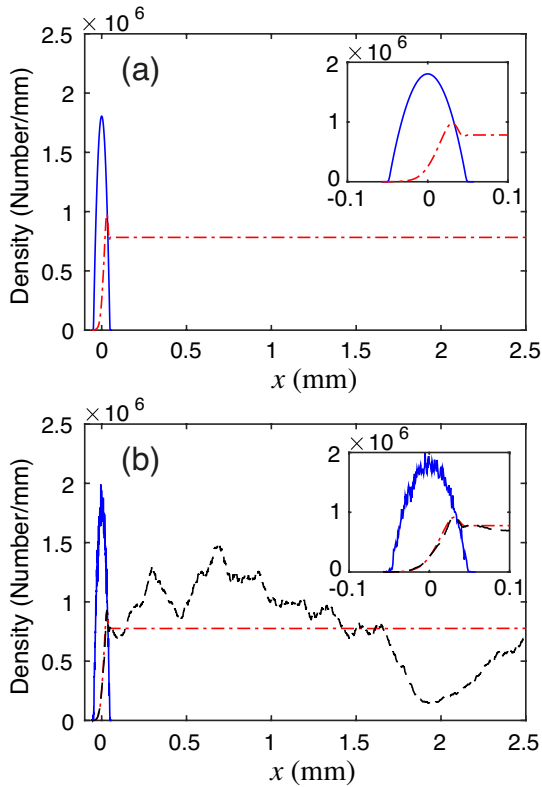}
	\caption{\label{fig:temptrace}(Color online) Typical classical field and atom laser beam profiles in steady state. (a) T = 0, (b) $T=144$ nK.  The instantaneous classical field density is shown in blue, whereas the instantaneous atom laser beam density is the black-dashed curve.  The red dot-dash curve shows the time-averaged atom laser beam density for comparison. The atom laser beam density has been scaled by a factor of $2\times10^3$ so that is visible on the axes as the trapped classical field. 
Insets: Detailed view of the classical field region. In (a), the instantaneous atom laser beam density and time-average atom laser beam density lie on top of one another. For all plots: $\mu = 102$ $\hbar \omega_x$, $g_{11}/\hbar\omega_x x_0 = 0.0166$ where $x_0 = (\hbar/m\omega_x)^{1/2}$, $\Omega = 1$ Hz, $\hbar \gamma = 0.0028$, $k_0 = 2.3\times10^{7}$ m$^{-1}$. .}
\end{figure}

We initialize our simulations with a $T=0$  condensate found from the solution of the time-independent Gross-Pitaevskii equation with the outcoupling turned off. The simulations are then run until we reach a steady state for the beam and condensate.  We calculate observables by averaging 500 time samples of 64--95 independent trajectories, which gives converged results with relatively small statistical noise. Fig.~\ref{fig:temptrace} shows examples of the instantaneous density of the trapped Bose gas and the atom laser beam once the system has reached steady state at (a) $T=0$, and (b) $T= 144$ nK. 

\section{Temporal coherence of the output beam}
\begin{figure}
	\centering
	\includegraphics[width=0.85\columnwidth]{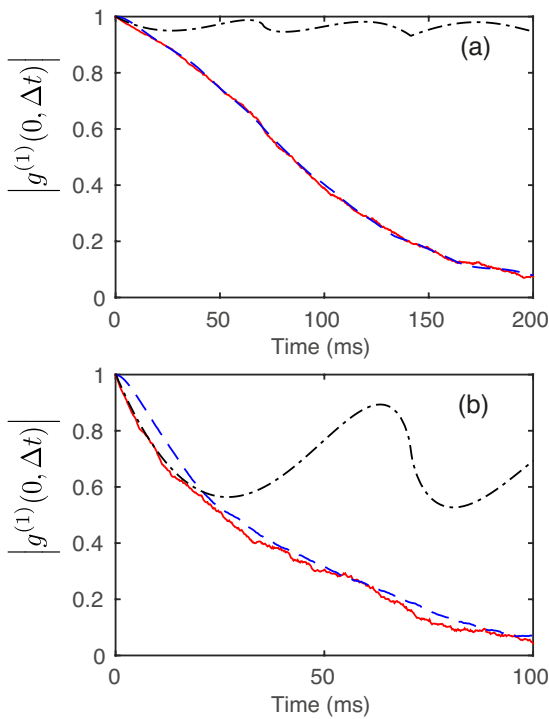}
	\caption{\label{fig:tempcomp} (Color online) First-order temporal coherence of the atom laser beam and condensate quantified by $g^{(1)}(x,\Delta t)$ for temperatures (a) $T=48$ nK with $N_0 = 1.1\times 10^5$ and (b) $T=384$ nK with $N_0 =0.72\times 10^5$. For both plots --- solid red line is the coherence of the condensate at $x=0$; dashed blue line: coherence of the atom laser beam for $x$ just outside the trapped gas; dash-dotted black line: estimate given by Bogoliubov approximation of Eq.~\eqref{eqn:g1_est}. All other parameters are the same as for Fig.~\ref{fig:temptrace}. }
\end{figure}

The phase coherence of the atom laser beam can be quantified by the normalised equal position first-order temporal correlation function \cite{Walls1994}
\begin{equation}
\label{eqn:g1}
 g^{(1)}(x,\Delta t)=\frac{\left\langle \phi^*(x,t)\phi(x,t+\Delta t)\right\rangle}{\sqrt{\left\langle |\phi(x,t)|^2\right\rangle \left\langle |\phi(x,t+\Delta t)|^2\right\rangle }} \, ,
\end{equation}
Figure \ref{fig:tempcomp} shows $ |g^{(1)}(x,\Delta t)|$ for two different temperatures.  Unsurprising, the coherence time decreases monotonically with increasing temperature for fixed $\mu$.  In this parameter regime we find that the atom laser coherence directly reflects the first-order temporal coherence of the classical field $\psi(x)$ at trap centre, which is plotted for comparison in Fig.~\ref{fig:tempcomp}

To gain physical insight into the decrease of phase coherence in the condensate  at finite temperature, we can approximate the evolution of the classical field by utilising the Bogoliubov approximation~\cite{stringari_review}.  Expressing the classical field as 
\begin{align}
{\psi}(x,t) &= e^{-i \mu t/\hbar}\left[\sqrt{N_0} \varphi_0(x)  \nonumber \right.\\
& \left.+\sum_{j>0} \left(u_j(x)e^{-i\epsilon_j t/\hbar}{b}_j + v^*_j(x)e^{i\epsilon_j t/\hbar} {b}^*_j\right) \right]\, , 
\end{align}
where $N_0$ is the condensate number, $\varphi_0(x) $ is the condensate wave function, and $u_j(x)$ and $v_j(x)$ are the 1D Bogoliubov quasiparticle modes with energies $\epsilon_j$ measured relative to the chemical potential~\cite{pitaevskii2003}.  An approximate form for the Bogoliubov modes in the 1D Thomas-Fermi limit can be written as \cite{Petrov2000} 
\begin{equation}
f^{\pm}_j(\tilde{x}) = \sqrt{j+\frac{1}{2}} \left[ \frac{2\mu}{\epsilon_j}(1-\tilde{x}^2) \right]^{\pm\frac{1}{2}} P_j(\tilde{x}),
\label{eqn:bog_soln}
\end{equation}
where the spatial dependence has been normalised to the Thomas-Fermi radius of the condensate $\tilde{x}=x/R_{\text{TF}}$ where $R_{\text{TF}} = (2 \mu/m \omega_x^2)^{1/2}$, and $P_j(\tilde{x})$ are Legendre polynomials.  The functions $f^{\pm}_j(\tilde{x})$ are related to the  Bogoliubov mode functions $u(\tilde{x})$ and $v(\tilde{x})$ via the relation $f^{\pm}_j(\tilde{x})=u(\tilde{x})\pm v(\tilde{x})$.  The Bogoliubov modes have energies $\epsilon_j=\hbar\omega_x \sqrt{j(j+1)/2}$ for positive integers $j$ \cite{Petrov2000} in the one-dimensional Thomas-Fermi limit.  

We can then  construct an estimate of the first-order correlation function $g^{(1)}(x,\Delta t)\propto \langle{\psi}^*(x,t){\psi}(x,t +\Delta t)\rangle$ by substituting the Bogoliubov expansion \eqref{eqn:bog_soln} and computing the sum numerically.  We examine the temporal coherence at the centre of the condensate where outcoupling is resonant.  We assume each Bogoliubov mode is thermally occupied with quasiparticles according to the equipartition distribution
\begin{equation}
N_j \equiv \langle {b}^*_j {b}_j\rangle = \frac{k_B T}{\epsilon_j},
\end{equation}
while the ground state has an occupation equal to the condensate number $N_0$ calculated from the one-body density matrix using the Penrose-Onsager criterion for condensation~\cite{Blakie2008}.  Assuming $N_j\gg 1$, we have
\begin{widetext}
\begin{equation}
\langle\psi^*(0,t)\psi(0,t+\Delta t)\rangle = N_0 |\varphi_0(0)|^2 
+ \sum_{j>0} \frac{k_BT}{\epsilon_j} \left(j+\frac{1}{2}\right)\frac{|P_j(0)|^2}{2}\left[\left(\frac{4\mu^2 + \epsilon_j^2}{2\mu\epsilon_j}\right)\cos\left(\epsilon_j \Delta t/\hbar\right)-2i\sin\left(\epsilon_j \Delta t/\hbar\right)\right],
\label{eqn:g1_est}
\end{equation}
\end{widetext}
The estimate given by Eq.~\eqref{eqn:g1_est} is also shown in Fig.~\ref{fig:tempcomp}, where the correlation function has been normalised to compare with the SPGPE results.  The estimate replicates the early decay of the numerical results, indicating that the mechanism for the loss of coherence observed in the condensate  is the dephasing of thermal fluctuations.  The revivals of coherence present in the Bogoliubov estimate are not present in the SPGPE result, as the noise term of the SPGPE causes a random drift in the phase of each of the modes and preventing the rephasing.

\section{Influence of the momentum kick on the linewidth}
We now investigate the effect of the magnitude of the momentum kick on the linewidth of the atom laser at finite temperature. At zero temperature the momentum transfer to the atom laser beam of $\hbar k_0$ would have no bearing on the temporal coherence and hence the linewidth of the atom laser.  At finite temperature we find that the temporal coherence of the trapped BEC is also independent of the value of $k_0$.  However, we find that this is not the case for the atom laser beam.  Instead, below a critical value of $k_0$, we find that for a given temperature the temporal coherence of the beam decays significantly faster than that of the trapped BEC. 

\begin{figure}
	\centering
	\includegraphics[width=0.85\columnwidth]{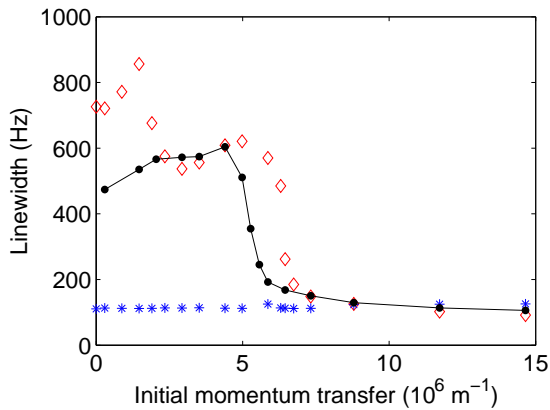}
	\caption{\label{fig:classicalcomp}
	(Color online) Linewidths of the condensate and the atom laser as a function of momentum kick for $T=384$ nK.  Blue asterisks: condensate linewidth at outcoupling position. Red diamonds: atom laser beam linewidth.  Black dots: Classical model of the linewidth  of Newtonian particles exiting the system experiencing the fluctuating classical field of the trapped BEC as described in the text. All other parameters are the same as for Fig.~\ref{fig:temptrace}.}
\end{figure}

Defining the power spectrum of a field $\chi(x_0,t)$  as
\begin{equation}
 P_\chi(x_0,\omega) = \left|\int dt e^{-i \omega t} \chi(x_0,t)\right|^2, 
\end{equation} 
we define the spectral linewidth of the field as 
\begin{eqnarray}
\Delta\omega_\chi(x_0)=\left[\int d\omega P_\chi(x_0,\omega) \omega^2 - \left(\int d\omega P_\chi(x_0,\omega) \omega\right)^2 \right]^{1/2} \, \nonumber  \\
\end{eqnarray}
This quantity is plotted for both the centre of the condensate  (blue asterisks) and the middle of the atom laser beam (red diamonds) in Fig.~\ref{fig:classicalcomp} for a temperature of $T=384$ nK.  It is clear that the atom laser beam exhibits a  dramatic increase in linewidth for momentum kicks below $k_0 =6\times10^6$ m$^{-1}$.

We have found that this excess linewidth is due to the thermal density fluctuations of the classical field that the atom laser beam experiences as it propagates through the trapped BEC.  If the outcoupling momentum $\hbar k$ is sufficiently large, the density of the trapped classical field is essentially frozen on the timescale required for particles to leave the region of the BEC, so the linewidth of the beam is determined by the spectral linewidth of the condensate  at the spatial position where the outcoupling is resonant. However, for smaller momenta, and therefore longer exit times, the outcoupled atoms experience significant spatio-temporal fluctuations in the effective spatial potential due to the classical field, resulting in greater dispersion in velocity of the outcoupled atoms and a larger linewidth.  A threshold momentum separating the different linewidth behaviour may be determined via a classical argument.  Particles travelling significantly faster than the thermal density fluctuations can `outrun' the fluctuations, and thus experience no velocity broadening.  Taking the sound speed $c_s=(g_{11}N_0 |\varphi(0)|^2/m)^{1/2}$  as an estimate for the characteristic speed of density fluctuations, we note that the threshold momentum for the excess linewidth is approximately twice this value, consistent with our argument.

We have implemented a simple numerical model  to confirm this explanation.  We have simulated the dynamics of an ensemble of classical atom laser beam particles placed at the classical outcoupling position with initial momentum $\hbar k_0$.  In a zero temperature system the effective potential experienced by the outcoupled atoms due to interactions with the BEC will be static.  Therefore, outcoupling the atom laser beam from the centre of the trap will result in a momentum increase of the output as the atoms ``roll down the hill'' so their output momentum is $\hbar k=\sqrt{(\hbar k_0)^2+2m\mu}$ \cite{Johnsson2007_1}, but there is no broadening of the initial distribution.  However, at finite temperature, on exiting the trapped BEC, the momentum distribution of the beam will be broadened due to interactions with the  fluctuating density of the classical field, and we quantify it by standard deviation of the output energy distribution.  The resulting linewidth is plotted in Fig.~\ref{fig:classicalcomp} in comparison with the results from our SPGPE model, where we can see they are in broad agreement.

As the excess linewidth of the atom laser beam is due to density fluctuations of the trapped BEC, the effect will disappear if there are no interactions between the trapped BEC and the atom laser beam.  We have verified this by setting $g_{12}=0$ in our simulations (while appropriately adjusting the detuning $\delta= \hbar k_0^2/2m-\mu/\hbar$ in order to maintain resonant outcoupling in the centre of the condensate.) In this case, the linewidth of the atom laser beam reverts to the linewidth of the condensate  for all values of the Raman kick momentum. It may be possible to exploit an inter-component Feshbach resonance to achieve this in an experiment, significantly reducing the linewidth. However, this may be technically challenging in practice. 

A potential solution to reduce the atom laser linewidth for a fixed temperature and momentum kick is to alter the detuning such that the resonant position is closer to the edge of the condensate and the exit time of the beam is shorter.  We do observe a monotonic decrease in the linewidth as the outcoupling position is moved towards $R_\text{TF}$, but it remains larger than that of the condensate.  Beyond $R_\text{TF}$, the atom laser linewidth approaches that of the thermal reservoir, as expected --- only thermal atoms are outcoupled. As a result, the  linewidth of the atom laser beam is at least that of a thermal state for all values of the two-photon Raman detuning.  We conclude that altering the detuning may decrease the overall atom laser linewidth, but cannot reduce it to that of the condensate.  Furthermore, any reduction in linewidth is offset by the reduction in flux (for a fixed $\Omega$) associated with moving the outcoupling position away from the centre of the trap.

Another strategy to reduce the excess linewidth observed for a given temperature and momentum kick may be to remove the waveguide and allow outcoupled atoms to instead fall under gravity in the radial direction, potentially reducing the exit time of the atom laser beam.  For the parameters used in this work, the gravitational acceleration results in an exit time equivalent to an initial momentum kick of $4.4\times10^6$ m$^{-1}$.  This value is below the threshold momentum, and so the linewidth of the atom laser beam will still be excessive. Additionally, removing the waveguide will degrade the transverse profile of the atom laser beam \cite{Guerin2006, Couvert2008}, as was seen in \cite{Coq2001,Busch2002, Kohl2005, Riou2006, Jeppesen2008}.

\section{Conclusions}
We have modelled the temporal coherence and linewidth of a continuous atom laser at finite temperature.   We find that when outcoupling with a large momentum kick, the first-order coherence of the atom laser beam mimics the first-order coherence of the trapped BEC which is determined by thermal fluctuations. However, below a threshold momentum kick related to the condensate speed of sound, there is a dramatic increase in the spectral linewidth of the atom laser beam. We have identified the origin of this as density fluctuations in the trapped BEC while the atom laser beam exits the condensate, further increasing the energy spread of the output beam.  Our results will influence the design and optimisation of continuous atom lasers at finite temperature in the future.

\begin{acknowledgements}
This research was supported by Australian Research Council Discovery Projects DP1094025 and DE130100575. ASB acknowledges the support of The New Zealand Marsden Fund and the Royal Society of New Zealand. MJD acknowledges the support of the JILA Visiting Fellows program.
\end{acknowledgements}

\bibliography{atomlaser_linewidth_bib}

\end{document}